\documentstyle[preprint,revtex]{aps}

\begin{document}
\hfill\vbox{\hbox{NUHEP-TH-93-30}\hbox{UCD-93-35}\hbox{December 1993}}\par
\thispagestyle{empty}
\begin{title}
\begin{center}
{\bf $B_c$ Meson  Productions Via Induced Gluon Fragmentation}
\end{center}
\end{title}
\author{Kingman Cheung}
\begin{instit}
Dept. of Physics \& Astronomy, Northwestern University, Evanston,
IL 60208\\
\end{instit}
\author{Tzu Chiang Yuan}
\begin{instit}
Davis Institute for High Energy Physics \\
Dept. of Physics, University of California, Davis, CA 95616 \\
\end{instit}

\begin{abstract}
\nonum

\section{Abstract}

Gluons cannot directly fragment into $B_c$ mesons until at order
$\alpha_s^3$.
However,  the Altarelli-Parisi evolution of
the $\bar b$-quark fragmentation functions
$D_{\bar b \to B_c}(z)$ and $D_{\bar b \to B_c^*}(z)$
{}from the heavy quark mass scales up to the collider energy scale $Q$
can induce the gluon fragmentation functions
$D_{g \to B_c}(z)$ and $D_{g \to B_c^*}(z)$, respectively,
at the order $\alpha_s^3 \, {\rm log}  (Q/m_{b,c})$
through the gluon splitting $g\to \bar b$.
We will determine these induced gluon fragmentation functions.
The $B_c$ meson  productions due to the induced gluon
fragmentation will be evaluated and compared
to the productions by the direct $\bar b$-quark fragmentation.
The contribution from the induced gluon fragmentation is found to be
a fairly large portion of the total $B_c$ and $B_c^*$ meson productions,
and therefore cannot be ignored.

\end{abstract}

\newpage

\section{INTRODUCTION}

The next and the last family of  $B$ mesons to be observed
will be the $B_c$ mesons made up of one charm quark and one bottom anti-quark.
The $B_c$ meson family
differs from the $J/\psi$ and  $\Upsilon$ families and from other $B$ mesons
because it is made up of a pair of heavy quark and anti-quark  of {\it
different} flavors and masses.  Unlike the other heavy-light mesons, a lot of
the dynamics of $B_c$ meson can be predicted reliably because it is within
the perturbative QCD regime like the other heavy quarkonium systems.
But it does not mean
that the $B_c$ mesons are not interesting sources of
new physics.  Instead, their
productions, spectroscopy, and decays will
offer great opportunities to examine various
heavy quark fragmentation models, heavy quark spin-flavor symmetry, different
quarkonium bound-state models, and properties of inclusive and exclusive
decay channels.

All the previous estimates \cite{bclep}
of $B_c$ meson production were based on the
perturbative calculation of $e^+e^- \to B_c b \bar c$, in which the ratio
$\sigma(B_c)/\sigma(b\bar b)$ was extracted and identified
as the fragmentation probability.
It was about $10^{-3}-10^{-4}$.  The uncertainty comes from the choices of the
quark masses and of the scale in the running strong coupling $\alpha_s$.
There was also an estimate \cite{bcmc} based on the
HERWIG monte carlo, and the above ratio is roughly
$1-2\times 10^{-3}$ at various colliders.
But there were no reliable estimates that can include higher order
corrections. This problem becomes more serious at hadronic colliders
because those terms consisting
of large logarithms of $(Q/m_{b,c})$  from
higher order corrections are needed to be summed in
order to get reliable predictions. Here $Q$ stands for the typical energy
scale in high energy hadronic collisions.

Recently, Braaten and Yuan \cite{gluon} pointed out that the
heavy quarkonium production in the  large transverse momentum region at
hadron colliders is dominated by fragmentation.
They also showed that
the fragmentation function $D_{g \to J/\psi}(z)$  for a gluon to split into
a $J/\psi$ with momentum fraction $z$
can be calculated within the framework of perturbative QCD.
Subsequently, the fragmentation functions for the following splitting processes
$c\to J/\psi$ \cite{psi}, $c\to$ polarized $J/\psi$
\cite{wise}, $\bar b \to B_c,B_c^*$ \cite{bc}, and $\bar b \to$ polarized
$B_c^*$ \cite{bcpol,chen}  were also evaluated.
Such developments are important because these
fragmentation  functions are the boundary conditions for the
Altarelli-Parisi evolution equations at the heavy quark mass scales.
Large logarithms can now be summed by solving the evolution equations.
The fragmentation functions are
process-independent and can be applied
not only to the $e^+e^-$ colliders but also to the hadronic environments.
The  work \cite{bckm} by one of us on the $B_c$ meson productions at
hadronic colliders was performed  based on the direct $\bar b \to B_c,B_c^*$
fragmentation functions \cite{bc}, and it was shown that a
sizeable  number of $B_c$ mesons should be seen
at the Tevatron in the very near future.

It was mentioned by Falk {\it et al.} in Ref.~\cite{wise} that as one
evolves from the charm quark mass scale up to the collider energy scale,
a gluon fragmentation function for  $g \to J/\psi$
can be induced by the splitting of $g \to c$ (or $\bar c$) followed by
$c$ (or $\bar c$) fragmenting into $J/\psi$,  and
the induced gluon fragmentation function can easily surpass
the direct gluon fragmentation.
This observation prompts us to examine the $B_c$ meson productions
due to the induced $g \to B_c,B_c^*$   fragmentation functions.
We will show despite the fact that
the induced gluon fragmentation functions are
at least an order of magnitude smaller than the corresponding direct
$\bar b$-quark fragmentation functions,  their impacts on
the $B_c$ meson productions
are nevertheless significant when compared to the
productions by the direct $\bar b\to B_c$ meson fragmentation.
The induced gluon contributions can account for about one-fifth of the
total production cross section with $p_T(B_c)>10$ GeV at the Tevatron,
and therefore cannot be ignored.

In the next section, we will determine the induced gluon fragmentation
functions. In section III, we will calculate the $p_T$ distributions and the
total cross sections for $B_c$ and $B_c^*$ meson productions due to
the induced gluon fragmentation at the Tevatron and the LHC,
and compare them with the productions by the
direct $\bar b$-quark fragmentation.
We will then conclude in section IV.

\section{THE INDUCED GLUON FRAGMENTATION FUNCTIONS}

Simply by using power counting, the direct
$g\to B_c,B_c^*$ fragmentation functions are
suppressed by an extra factor
of $\alpha_s$ relative to the
direct $\bar b \to B_c,B_c^*$ fragmentation functions.
However, in the leading logarithmic approximation,
the induced gluon fragmentation functions
$D_{g\to B_c,B_c^*}(z,\mu)$ are of order $\alpha_s^3 \log(\mu/m_{b,c})$
at a renormalization scale $\mu$ \cite{wise}.
At large enough scale $\mu$, the logarithmic factor $\log(\mu/m_{b,c})$
may compensate for the extra power of $\alpha_s$ such that the induced
$D_{g\to B_c,B_c^*}(z,\mu)$ are effectively of order $\alpha_s^2$, just
the same as the direct $\bar b$-quark fragmentation functions
$D_{\bar b \to B_c,B_c^*}(z)$.
For this reason, we will ignore the direct $g \to B_c,B_c^*$
contributions in what follows.
We will also ignore the induced gluon fragmentation
associated with $g \to c$ followed by $c$ fragmenting
into $B_c$ and $B_c^*$ since the fragmentation probabilities for
$c \to B_c,B_c^*$ are two orders of magnitude smaller than those for
$\bar b \to B_c,B_c^*$ \cite{bc}. The
Altarelli-Parisi evolution equations for
$D_{\bar b\to B_c}(z)$ and $D_{g\to B_c}(z)$ are given by
\begin{equation}
\label{Db}
\mu \frac{\partial}{\partial \mu} D_{\bar b\to B_c}(z,\mu) =
\int_z^1 \frac{dy}{y}
P_{\bar b\to \bar b}(z/y,\mu)\; D_{\bar b \to B_c}(y,\mu) +
\int_z^1 \frac{dy}{y} P_{\bar b\to g}(z/y,\mu)\; D_{g \to B_c}(y,\mu) \,,
\end{equation}
\begin{equation}
\label{Dg}
\mu \frac{\partial}{\partial \mu} D_{g\to B_c}(z,\mu) = \int_z^1 \frac{dy}{y}
P_{g \to \bar b}(z/y,\mu)\; D_{\bar b \to B_c}(y,\mu) +
\int_z^1 \frac{dy}{y} P_{g \to g}(z/y,\mu)\; D_{g \to B_c}(y,\mu) \,.
\end{equation}
A similar set of equations associated with the $B_c^*$ case can also be
written down but will be omitted here. If the fragmenting
gluon and $\bar b$-quark have energies
much greater than the heavy quark masses,
the Altarelli-Parisi splitting functions
can be taken to be those of the light quarks, i.e., to leading order
in $\alpha_s$,
\begin{eqnarray}
\label{Pbb}
P_{\bar b\to\bar b}(x,\mu) &=&
\frac{4\alpha_s(\mu)}{3\pi} \left( \frac{1+x^2}{1-x} \right )_+ \,, \\
\label{Pbg}
P_{\bar b \to g}(x,\mu) &=& \frac{4\alpha_s(\mu)}{3\pi} \left(
\frac{1+(1-x)^2}{x} \right ) \,, \\
\label{Pgb}
P_{g\to\bar b}(x,\mu) &=& \frac{\alpha_s(\mu)}{2\pi} \left( x^2+(1-x)^2 \right
)
\,, \\
\label{Pgg}
P_{g \to g}(x,\mu) &=& \frac{6 \alpha_s(\mu)}{\pi} \left(
{x \over (1-x)_+} + {1-x \over x} + x(1-x) + \left(
\frac{11}{12}-\frac{n_f}{18} \right ) \delta (1-x) \right) \, ,
\end{eqnarray}
where $n_f$ is the number of active flavors below the scale $\mu$.
Since the energy threshold of producing a $B_c$ or $B_c^*$ meson
{}from a gluon is $2(m_b+m_c)$ in the non-relativistic approximation, we assume
\begin{equation}
D_{g\to B_c}(z, \mu) = D_{g\to B^*_c}(z, \mu) =
0 \qquad {\rm for} \qquad \mu \le 2(m_b+m_c) \, .
\label{dg0}
\end{equation}
The $D_{\bar b\to B_c}(z,\mu_0)$ and $D_{\bar b\to B_c^*}(z,\mu_0)$
at the initial scale $\mu_0=m_b+2m_c$ are given by \cite{bc}
\begin{eqnarray}
D_{\bar b\rightarrow B_c}(z,\mu_0) & = &
\frac{2\alpha_s(2m_c)^2 |R(0)|^2}
{81\pi m_c^3}\; \frac{rz(1-z)^6}{(1-(1-r)z)^6} \nonumber \\
\label{dz1}
&\times & \Biggl[ 6 - 18(1-2r)z + (21 -74r+68r^2) z^2 \\
 && -2(1-r)(6-19r+18r^2)z^3  + 3(1-r)^2(1-2r+2r^2)z^4 \Biggr]\,,  \nonumber
\end{eqnarray}
for the $B_c$ state, and
\begin{eqnarray}
D_{\bar b\rightarrow B_c^*}(z,\mu_0) & = &
\frac{2\alpha_s(2m_c)^2 |R(0)|^2}
{27\pi m_c^3}\; \frac{rz(1-z)^6}{(1-(1-r)z)^6} \nonumber \\
\label{dz2}
&\times & \Biggl[ 2 - 2(3-2r)z + 3(3 - 2r+ 4r^2) z^2 \\
& &   -2(1-r)(4-r +2r^2)z^3  + (1-r)^2(3-2r+2r^2)z^4 \Biggr] \,,  \nonumber
\end{eqnarray}
for the $B_c^*$ state.
$R(0)$ is the radial wave function at the origin
with the radial quantum number suppressed and $r=m_c/(m_b+m_c)$.
We can now evolve the functions $D_{\bar b\to B_c,B_c^*}(z)$ and
$D_{g\to B_c,B_c^*}(z)$ from $\mu_0$ to a higher scale by solving
the evolution equations (\ref{Db}) -- (\ref{Pgg}) with
Eqns.(\ref{dg0}) -- (\ref{dz2}) as the boundary conditions.
Despite the off-diagonal
terms
$P_{\bar b \to g}\otimes D_{g\to B_c}$ and
$P_{\bar b \to g}\otimes D_{g\to B^*_c}$
are included here,
the evolved $D_{\bar b\to B_c}(z)$ and $D_{\bar b\to B_c^*}(z)$ have
no noticeable difference from those shown in Ref.~\cite{bc} and will not be
repeated.  Because the splitting function
$P_{\bar b\to \bar b}$ is a total plus--function,
the $\bar b\to B_c,\,B_c^*$ fragmentation probabilities remain the same
over all scales except for  a small correction from
the off-diagonal terms.
The fragmentation probability for $\bar b\to B_c$ is about
$3.93(4.05)\times 10^{-4}$, and that for $\bar b\to B_c^*$ is about
$5.53(5.70)\times 10^{-4}$ at the initial scale
$\mu_0$ = 7.9 GeV
(the very large scale of 0.8 TeV).
We chose $m_c=1.5$ GeV and $m_b=4.9$ GeV.
The induced gluon fragmentation functions
$D_{g\to B_c}(z,\mu)$ and $D_{g\to B_c^*}(z,\mu)$ at various scales $\mu$ are
shown in Fig.~\ref{fig-Dg}.
The fragmentation probabilities
$\int dz D_{g\to B_c}(z,\mu)$ and $\int dz D_{g\to B_c^*}(z,\mu)$
at different scales $\mu$ are also listed in Table~\ref{table1}.  From this
Table,
one can see that the fragmentation probabilities for $g\to B_c,\,B_c^*$
increase rapidly as the scale $\mu$ goes  up. The
induced gluon fragmentation probabilities are in general only a few \% to
at most 10\% of the direct  $\bar b\to B_c,B_c^*$
fragmentation probabilities for $\mu$ ranging
{}from a few ten GeV's  up to 0.8 TeV.
Hence, for the range of scales that is relevant to the $B_c$ meson
productions,   the induced gluon fragmentation functions
$D_{g \to B_c,B_c^*} (z,\, \mu)$  are not important relative to the
direct $\bar b$-quark fragmentation functions
$D_{\bar b \to B_c,B_c^*}(z,\mu)$.
The reason why the induced $g\to B_c,B_c^*$ fragmentation is important
to the $B_c$ meson productions is not because of the induced
gluon fragmentation functions themselves, which we have shown to be small,
but because the subprocess cross sections to produce
gluon are  much larger than those to produce
$\bar b$-quark.
This is the subject that we are now turn to in the next section.

\newpage

\section{$B_c$ MESON PRODUCTIONS AT HADRON COLLIDERS}

The induced gluon fragmentation functions
$D_{g\to B_c}(z)$ and $D_{g\to B_c^*}(z)$
are convoluted with the
inclusive gluon production subprocesses to obtain the cross sections for
$B_c$ and $B_c^*$ meson productions, respectively.
The differential cross section of  the $B_c$ meson production
at a transverse momentum $p_T$ is given by
\begin{eqnarray}
d\sigma\left(B_c   (p_T) \right)
&=& \sum_{ij} \int dz \int dx_1 \int dx_2
\;f_{i/p}(x_1,\mu) f_{j/p}(x_2,\mu) \nonumber \\
\label{pt}
&&\quad \times
d \hat \sigma ( ij \to g(p_T/z) X,\,\mu) \; D_{g\to B_c}(z,\,\mu)\,,
\end{eqnarray}
where the summation is over all possible subprocesses:
$gg\to gg$, $gq(\bar q) \to gq(\bar q)$, and $q\bar q \to gg$.
Formulas for the subprocess cross sections $d \hat \sigma$ at the lowest order
 can be found in standard textbooks.
Similar equation can be written down for the case of $B_c^*$ and
will be omitted.
We use the parameterization of HMRS \cite{HMRS} for the parton
distribution functions $f_{i/p}(x)$.  The running strong coupling
$\alpha_s$ is evaluated at 1-loop by evolving from the
experimental value $\alpha_s(m_Z)=0.12$, and is given by
\begin{equation}
\alpha_s(Q) = \frac{\alpha_s(m_Z)}{1+ 8\pi b_0\alpha_s(m_Z)\log(Q/m_Z)}\,,
\end{equation}
where $b_0=(33-2n_f)/48\pi^2$ and $n_f$ is the number of active flavors below
the scale $Q$.
The factorization scale $\mu$  in Eqn.(\ref{pt})
is chosen to  be the transverse momentum $p_{Tg}$ of the fragmenting gluon.

The transverse momentum distributions for the $B_c$ and
$B_c^*$ mesons with $p_T(B_c,B_c^*)>10$~GeV and $|y(B_c,B_c^*)|<1$
at the Tevatron are shown in Fig.~\ref{fig-pt} for both the induced
gluon fragmentation and the direct $\bar b$-quark fragmentation.
The corresponding curves at the LHC with $p_T(B_c,B_c^*)>20$~GeV and
$|y(B_c,B_c^*)|<2.5$ are shown in Fig.~\ref{fig-lhc}.
The integrated cross sections
$\sigma(p_T(B_c)> p_T^{\rm min}(B_c))$ and
$\sigma(p_T(B_c^*)> p_T^{\rm min}(B_c^*))$
are shown in Fig.~\ref{fig-inte}.
Although the induced gluon fragmentation probabilities are only
a few \% of the  corresponding
direct $\bar b$-quark fragmentation probabilities, the
amplitude squared of the most important
subprocess $gg\to gg$ is more than an order of magnitude larger than that
of  the subprocess $gg \to b\bar b$.  It means that the large cross section
of the subprocess $gg\to gg$
makes up the smallness of the induced gluon fragmentation functions
such that the
resultant $B_c$ meson production cross sections
due to the induced gluon fragmentation
are not insignificant.  As shown in Fig.~\ref{fig-inte},
the contributions from the induced $g\to B_c,\, B_c^*$ fragmentation
are almost 30\% (50\%)  of the
corresponding contributions from  the direct $\bar b\to B_c,\, B_c^*$
fragmentation with $p_T(B_c,\,B^*_c) >10 (30)$~GeV.
Actually, the contribution from the induced gluon fragmentation gets closer to
the contribution from the direct $\bar b$-quark fragmentation at higher
$p_T(B_c, \,B^*_c)$
because the induced $g\to B_c, \,B_c^*$ fragmentation has
a larger probability as the scale $\mu=p_{Tg}$ increases
(illustrated  clearly in both Table~\ref{table1} and Fig.~\ref{fig-Dg}).
The situation at the LHC is roughly the same,
as shown in Figs.~\ref{fig-lhc} and \ref{fig-inte}.
This is due to the fact that we are comparing mainly the subprocess
$gg\to gg$ followed by the induced $g\to B_c,B_c^*$
fragmentation and the subprocess
$gg\to b\bar b$ followed by the direct $\bar b\to B_c,B_c^*$
fragmentation. The comparison is therefore not affected by the enormous $gg$
luminosity at the LHC compared with the Tevatron.

So far we have only presented the results of the $1S$ states.
All the excited $B_c$ states will
eventually decay into the  ground state $1^1S_0$ via emissions of pions and
photons, and so they all contribute to the inclusive $B_c$ meson  production.
The contribution from the first radial excited states $2S$ can be obtained
simply by multiplying the $1S$ results by the ratio
$|R_{2S}(0)|^2/|R_{1S}(0)|^2\approx 0.6$.  At the Tevatron, for an integrated
luminosity of 25~pb$^{-1}$ there are about
16,000 $B_c$ mesons with $p_T(B_c)>10$~GeV \cite{bckm}
by the direct $\bar b$-quark  fragmentation including all the
$1S$ and $2S$ states;  while there should be about
4,300 $B_c$ mesons produced by the induced gluon fragmentation with the same
integrated luminosity.  Thus we have shown that the induced gluon
fragmentation increases appreciably the total $B_c$ meson production
that was reported in Ref.~\cite{bckm},
where only the direct $\bar b$-quark fragmentation was considered.

\section{CONCLUSIONS}

We have determined the induced gluon fragmentation functions $D_{g\to B_c}(z)$
and $D_{g\to B_c^*}(z)$ for gluon to fragment into the two S-wave
heavy meson states, $B_c$ and $B_c^*$, composed of a charm quark and
a bottom anti-quark. They are only a few \% of the corresponding direct
$\bar b$-quark fragmentation functions.
%
%Despite their
%smallness in sizes, their effects are magnified at the high energy hadronic
%machines when they are convoluted with the large parton cross section
%of $gg \to gg$.
%
But due to the large parton-level cross sections of the subprocesses,
the $B_c$ and $B_c^*$ meson productions by the induced gluon fragmentation
are found to be significant and comparable to the
productions by the direct $\bar b$-quark fragmentation.
Hence,  we have shown that the induced gluon
fragmentation increases appreciably the total $B_c$ meson production.
For an integrated luminosity of 25 pb$^{-1}$ at the Tevatron,
there are totally about 20,000 $B_c$ mesons with $p_T(B_c)>10$~GeV
{}from both the induced gluon and direct $\bar b$-quark fragmentation,
of which 4,300 come from the induced gluon contribution.
Taking into account of the branching ratio ($\approx
0.2$\%) of the most spectacular signature of three charged leptons from the
same secondary vertex in the decay $B_c \to J/\psi \bar \ell' \nu_{l'} \to
(\ell\bar\ell)\bar \ell' \nu_{l'}$, there should be a sizeable number of
$B_c$ mesons that can be observed very soon at the Tevatron.

The smallness of the induced gluon fragmentation functions
also remind us that the contribution from
direct gluon fragmentation, which we have ignored in this analysis, might
be important as well. It might be interesting to calculate the ${\cal
O}(\alpha_s^3)$ direct gluon fragmentation functions. They, of course,
can only increase the production rates that we have already evaluated.
Extensions of the above results to the P-wave and D-wave $B_c$ mesons
are also important. Since the annihilation channels of these excited
$B_c$ meson states are highly suppressed, once produced by fragmentation
they will eventually decay
into the S-wave ground state $B_c$. Hence, they all contribute to the inclusive
$B_c$ meson production.  Fragmentation functions for
$\bar b \to$ P-wave $B_c$ mesons have now been calculated \cite{qpwave}.

Similar conclusions like those obtained in this paper can be
made for the $J/\psi$ (or $\psi'$) production at hadron colliders.
The $g\to J/\psi$ fragmentation induced by $g \to c$ or $\bar c$
followed by $c$ or $\bar c \to J/\psi$ (or $\psi'$)
is about two orders of magnitude smaller than the
direct $c$ or $\bar c \to J/\psi$ (or $\psi'$) fragmentation \cite{psi},
but should be comparable to or even larger than
the direct $g\to J/\psi$ (or $\psi'$) fragmentation \cite{gluon},
as already pointed out in Ref.~\cite{wise}.
Also, the $J/\psi$ (or $\psi'$) production by the induced gluon
fragmentation should be a significant fraction of the
production by the direct $c$ (or $\bar c$)
$\to J/\psi$ (or $\psi'$) fragmentation,
mainly because the subprocess $gg\to gg$ has a  much larger
cross section than the subprocesses $gg\to c\bar c$ and $c(\bar c) g \to c
(\bar c) g$.   However, the dominant production mechanism
of the P-wave charmonium states $h_c$ and $\chi_{cJ}(J=0,1,2)$ at large
transverse momentum are due to the direct gluon and direct $c$-quark
(or $\bar c$-quark) fragmentations,  since they both are
of order $\alpha_s^2$ \cite{wise,qpwave,gpwave}!
Radiative decays of these P-wave states into $J/\psi$ can contribute to
the inclusive $J/\psi$ production at hadron colliders.
Preliminary results on direct charmonium production via fragmentation
have been reported in Ref.\cite{charmonium}.

\acknowledgements

We are grateful to Eric Braaten and  Mark Wise  for useful discussions.
This work was supported by the U.~S. Department of Energy, Division of
High Energy Physics, under
Grants DE-FG02-91-ER40684 and DE-FG03-91ER40674 and
by Texas National Research Laboratory Grant RGFY93-330.

%-----------------------------------

%%%%%%%%%%%%%%%%%%%%%%%%%%%%%%%%%%%%%%%%%%%%%%%%%%%%%%%%%%%%%%%%%%%%%%%%%
\begin{table}
\caption{\label{table1}
The fragmentation probabilities $\int dz D_{g\to B_c}(z,\,\mu)$ and
$\int dz D_{g\to B_c^*}(z,\,\mu)$ at various scale $\mu$.
Since the induced gluon fragmentation functions diverge at $z=0$, we
only integrate from $z=0.01$ to $z=1$.
The direct $\bar b\to B_c$ and $\bar b\to
B_c^*$ fragmentation probabilities are $3.93\times 10^{-4}$ and
$5.53\times 10^{-4}$ respectively at the initial scale 7.9 GeV.
}
\begin{tabular}{ccc}
Scale $\mu$ & $\int dz D_{g\to B_c}(z,\,\mu)$ &$\int dz D_{g\to
B_c^*}(z,\,\mu)$
\\
\hline
15 &        $1.1 \times 10^{-6}$     &    $1.6 \times 10^{-6}$ \\
20 &        $3.4 \times 10^{-6}$     &    $4.8 \times 10^{-6}$ \\
30 &        $7.0 \times 10^{-6}$     &    $1.0 \times 10^{-5}$ \\
40 &        $9.8 \times 10^{-6}$     &    $1.4 \times 10^{-5}$ \\
60 &        $1.4 \times 10^{-5}$     &    $2.0 \times 10^{-5}$ \\
80 &        $1.7 \times 10^{-5}$     &    $2.4 \times 10^{-5}$ \\
100 &       $1.9 \times 10^{-5}$     &    $2.8 \times 10^{-5}$ \\
200 &       $2.7 \times 10^{-5}$     &    $3.9 \times 10^{-5}$ \\
400 &       $3.5 \times 10^{-5}$     &    $5.0 \times 10^{-5}$ \\
600 &       $3.9 \times 10^{-5}$     &    $5.7 \times 10^{-5}$ \\
800 &       $4.3 \times 10^{-5}$     &    $6.1 \times 10^{-5}$ \\
\end{tabular}
\end{table}

%%%%%%%%%%%%%%%%%%%%%%%%%%%%%%%%%%%%%%%%%%%%%%%%%%%%%%%%%%%%%%%%%%%5

\figure{
\label{fig-Dg}
The induced gluon fragmentation functions $D_{g\to B_c}(z,\mu)$
and $D_{g\to B_c^*}(z,\mu)$ at various scales $\mu=20,\,50,$ and $100$ GeV.
For each $\mu$  the lower (upper) curve is for $B_c$ ($B_c^*$).}

\figure{
\label{fig-pt}
The differential cross sections $d\sigma/dp_T(B_c)$
($d\sigma/dp_T(B_c^*)$) versus
the transverse momentum
$p_T(B_c)$ ($p_T(B_c^*)$)
of the $B_c$ ($B_c^*$) meson respectively at the Tevatron,
due to the direct $\bar b$-quark fragmentation  (solid)
and the induced gluon fragmentation (dashed).}

\figure{
\label{fig-lhc}
Same as Fig.~\ref{fig-pt} but at the LHC energy.}

\figure{
\label{fig-inte}
The integrated cross sections
$\sigma(p_T(B_c)>p_T^{\rm min}(B_c))$ (lower) and
$\sigma(p_T(B_c^*)>p_T^{\rm min}(B_c^*))$ (upper)  versus the
transverse momentum cut on the $B_c$ and $B_c^*$ mesons at the Tevatron and
the LHC.  Both contributions from the induced gluon fragmentation (dashed) and
direct $\bar b$-quark fragmentation (solid) are shown. }

\end{document}